\documentclass[conference]{IEEEtran}

\usepackage{graphicx}
\usepackage{amsfonts}
\usepackage{amsmath}
\usepackage{amssymb}
\usepackage{amsthm}
\usepackage{subfig}
\usepackage[font=small]{caption}
\usepackage{stfloats}
\usepackage{amsthm}
\usepackage{relsize}
\usepackage{rotating}
\usepackage{pgfplots}
\usepackage{siunitx}
\usepackage{cite}
\usepackage{mathtools}
\usepackage{algorithm}
\usepackage{algorithmic}
\usepackage{dsfont}

\usepackage{hyperref}
\usepackage{fancyhdr}

\DeclareMathOperator*{\argmax}{argmax}
\DeclareMathOperator*{\argmin}{argmin}

\newcommand*{\argminl}{\argmin\limits}
\newcommand*{\argmaxl}{\argmax\limits}

\begin{document}

\title{Serial Interference Cancellation for Improving uplink in LoRa-like Networks} 

\author{
{Angesom Ataklity TESFAY\textsuperscript{1}, Eric Pierre SIMON\textsuperscript{1}, Guillaume FERR\'E\textsuperscript{2} and Laurent CLAVIER\textsuperscript{1,3}} \\
\small
\textsuperscript{1}Univ. Lille, CNRS, UMR 8520 - IEMN, F-59000, Lille, France (e-mail: firstname.name@univ-lille.fr)\\
\textsuperscript{2}Univ. Bordeaux, CNRS, Bordeaux INP, IMS, UMR 5218, F-33400, Talence, France (e-mail: firstname.name@ims-bordeaux.fr)\\
\textsuperscript{3}IMT Lille Douai, France (e-mail: firstname.name@imt-lille-douai.fr)}

\maketitle
\thispagestyle{fancy}
\pagestyle{fancy}

\begin{abstract}
In this paper, we present a new receiver design, which significantly improves performance in the Internet of Things networks such as LoRa, i.e., having a chirp spread spectrum modulation. The proposed receiver is able to demodulate multiple users simultaneously transmitted over the same frequency channel with the same spreading factor. From a non-orthogonal multiple access point of view, it is based on the power domain and uses serial interference cancellation. Simulation results show that the receiver allows a significant increase in the number of connected devices in the network.
\end{abstract}


\IEEEpeerreviewmaketitle

\DeclareRobustCommand{\lc}{\raisebox{2pt}{\tikz{\draw[black,solid,line width=1.1pt](0,0) -- (7mm,0);}}}
\DeclareRobustCommand{\lt}{\raisebox{2pt}{\tikz{\draw[black,dashed,line width=1.1pt](0,0) -- (7mm,0);}}}
\DeclareRobustCommand{\lp}{\raisebox{2pt}{\tikz{\draw[black,dotted,line width=1.1pt](0,0) -- (7mm,0);}}}
\DeclareRobustCommand{\lpt}{\raisebox{2pt}{\tikz{\draw[black,dash pattern={on 7pt off 2pt on 1pt off 3pt},line width=1.1pt](0,0) -- (7mm,0);}}}

\section{Introduction}
Internet of Things (IoT) offers interconnection between a large number of physical objects
used for different applications such as environmental monitoring, smart farming and metering. The number of devices in the IoT network is expected to increase exponentially \cite{iot,iot2} in the coming years. 
Over recent years, Low Power Wide Area Networks (LPWAN) technologies, like LoRa and SigFox in unlicensed bands, 
are proposed to face this growth.  
LoRa employs a Chirp-Spread Spectrum (CSS) Modulation technique to encode information \cite{lora-Patent,lora}. Several Spreading Factors (SF) and bandwidth settings enable quasi-orthogonal transmissions for LoRa end-devices. 
However, when two or more end-devices transmit simultaneously on the same band and with the same SF, a collision occurs at the receiver.
This collision generally results in the loss of all colliding packets, but sometimes 
the first arrived signal might 
be fully decoded due to capture effect.

The authors in \cite{Analysis_lora} examines the interfering signal  
and analytically derives the performance of LoRa under same SF interference. Information of interferers is lost in the process. Recent work \cite{decode-lora} proposes an algorithm to decode the correct symbols of synchronized or slightly desynchronized signals using their timing information. However, ensuring synchronization is complex in the up-link of long range networks.  
 
Recently, the authors in \cite{enhanced-receiver} proposed a receiver which can decode two signals received at the same time, with the same spreading factor using serial interference cancellation (SIC). Asynchronicity is processed by estimating the time shift between the two received signals. However, the authors presumed that the signal with highest power was received first and this is not necessarily the case. The extension of this work, which is able to process multiple simultaneously  received  signals with the same SF is presented in \cite{enhanced-receiver-all}. However, the power ratio between two successive received signals is assumed to be constant.

The work in \cite{lora-like}, closely related to ours, proposes SIC to decode information from all users. A synchronized reception is considered, which is not realistic in the uplink. 

In this work, we propose a SIC technique to enable a receiver to decode  simultaneously multiple signals. We extend the work in \cite{lora-like} to a more general case where transmitters are asynchronous. 
Our main contributions are to develop a complete receiver structure, including the detection of packets, channel estimation, detection of symbols and interference cancellation. We analyse the performance in the presence of multiple interfering nodes and show that this approach can significantly improve the scalability of the network. This work is divided into five sections. Section II describes some key features of LoRa and introduces the system model. Section III presents the SIC algorithm while section IV discusses the simulation results. Conclusions are drawn in section V.
\setcounter{equation}{3}
\begin{figure*}
\hspace{0.1cm}\rule{17.75cm}{0.005cm}
\begin{align} \label{eq:3}
s_{p,1}^{(j,i)}(t)= \begin{cases}
&\exp\bigg(2\jmath\pi\Big(\frac{B}{2T_s}t^2 + \big(\frac{m_{p,1}^{(i)}-B\Delta^{(j,i)}}{T_s}\big)t + \phi_{p,1}^{(j,i)}\Big)\bigg),\\
& \hspace{5cm}  t\in A_{p,1}=\big [\frac{-T_s}{2} + \Delta^{(j,i)}, \frac{T_s}{2} + \Delta^{(j,i)} - \tau_{p,1}^{(i)}\big[,\\
&\exp\bigg(2\jmath\pi\Big(\frac{B}{2T_s}t^2 + \big(\frac{m_{p,1}^{(i)}-B\Delta^{(j,i)}}{T_s}\big)t + \phi_{p,1}^{(j,i)} - B(t-\Delta^{(j,i)})\Big)\bigg),\\
& \hspace{5cm} t\in B_{p,1} =\big [ \frac{T_s}{2} + \Delta^{(j,i)} - \tau_{p,1}^{(i)}, \frac{T_s}{2} \big [.
\end{cases}
\end{align}
\begin{align}\label{eq:4}
s_{p,2}^{(j,i)}(t)=\begin{cases}
&\exp\bigg({ 2\jmath\pi\Big(\frac{B}{2T_s}t^2 + \big(\frac{m^{(i)}_{p,2} + B(T_s - \Delta^{(j,i)})}{T_s}\big)t + \phi_{p,2}^{(j,i)}\Big)\bigg) }, \\
& \hspace{5cm}  t\in B_{p,2}=\big [ \frac{-T_s}{2},  \frac{-T_s}{2} + \Delta ^{(j,i)} - \tau_{p,2}^{(i)}\big [,\\
&\exp\bigg(2\jmath\pi\Big(\frac{B}{2T_s}t^2 + \big(\frac{m_{p,2}^{(i)}+B(T_s - \Delta^{(j,i)})}{T_s}\big)t + \phi_{p,2}^{(j,i)} - B(t+T_s-\Delta^{(j,i)})\Big)\bigg),\\
& \hspace{5cm} t\in A_{p,2} =\big [ \frac{-T_s}{2} + \Delta ^{(j,i)} - \tau^{(i)}_{p,2}, \frac{-T_s}{2} + \Delta^{(j,i)}\big [.
\end{cases}
\end{align}
\begin{equation*}
 \text{ where } \; \;\; \;\; \phi_{p,1}^{(j,i)}=\frac{B(\Delta^{(j,i)})^2 - 2m_{p,1}^{(i)}\Delta^{(j,i)} }{2T_s},
  \; \;\text{and} \; \;
\phi_{p,2}^{(j,i)}=\frac{B(T_s - \Delta ^{(j,i)})^2 + 2m_{p,2}^{(i)}(T_s - \Delta^{(j,i)}) }{2T_s}.
\end{equation*}
\hspace{0.1cm}\rule{17.75cm}{0.005cm}
\end{figure*}
\section{System Model}\label{sec:2}

The LoRa signal model is an $M$-ary digital modulation based on frequency shift chirp spread spectrum. The $M$ possible waveforms are chirps  modulated over the frequency band $B$ with $M$ different initial frequencies. A modulating symbol carries $\textrm{SF} = \log_2(M)$ bits, with SF  ranging from 7 to 12\footnote{$\textrm{SF}=6$ also exists but the modulation scheme is modified.}.
The symbol duration is $T_s = MT$, where $T = 1/B$ is the sampling period.
Each modulated chirp is based on a raw chirp, whose instantaneous frequency is given by 
$f(t)=\frac{B}{T_s}t$. Recall that the instantaneous phase ($\vartheta(t)$: $f(t)=\frac{1}{2\pi}\frac{d\vartheta(t)}{dt}$) is the integral of $f(t)$, which yields the following baseband expression for the raw chirp:
\setcounter{equation}{0}
\begin{equation}
c(t)= \exp\Big(2\jmath\pi\frac{B}{T_s}\frac{t^2}{2}\Big), \;  t\in\Big[ -\frac{T_s}{2}, \frac{T_s}{2}\Big[.
\end{equation}
\normalsize

\begin{figure}
    \centering
    \includegraphics[scale=0.55]{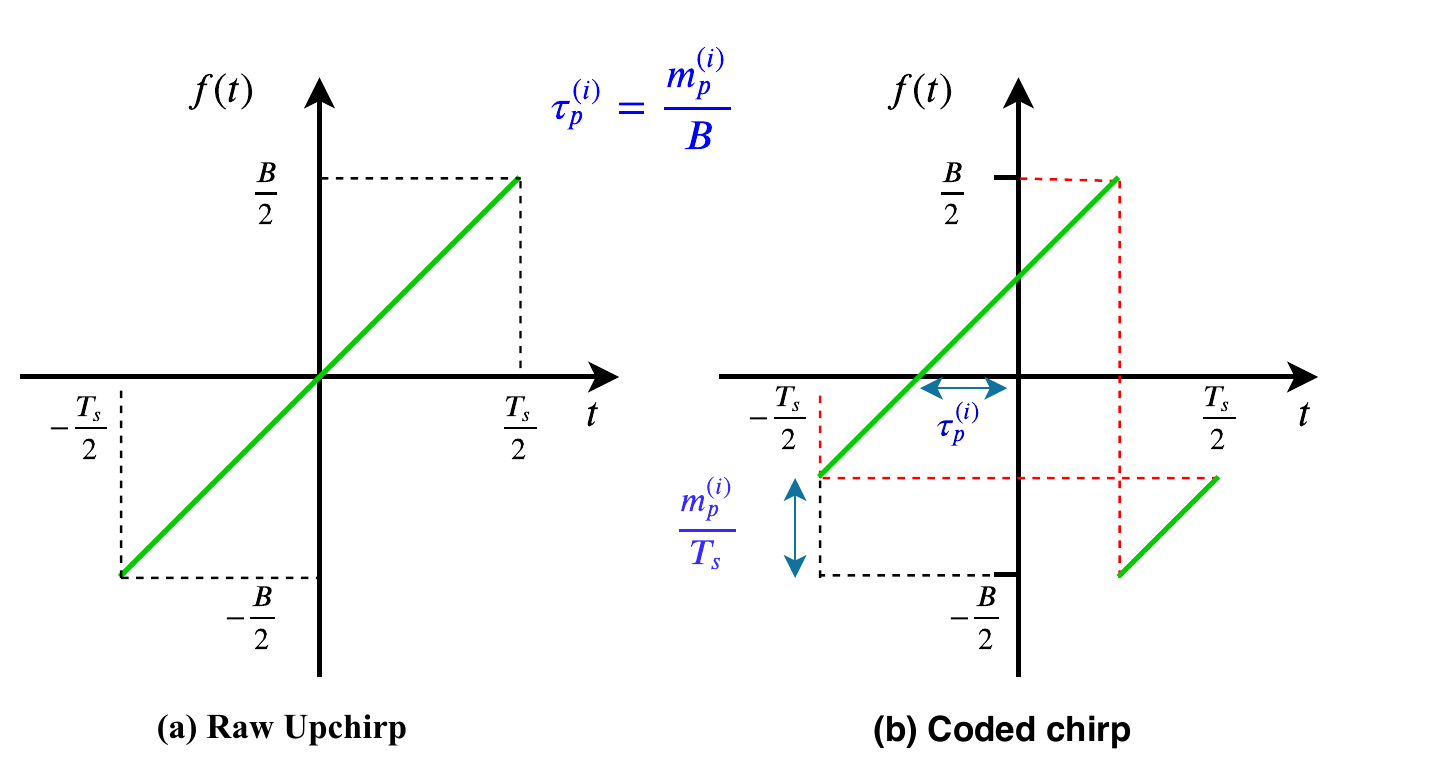}
    \caption{Raw up-chirp and Coded chirp associated with $m^{(i)}_p$.}
    \label{fig:1}
\end{figure}

The transmitted  modulating symbol of the $i$th user at time $pT_s,\;p= 0,\cdots,P-1$, with $P$ the number of  symbols transmitted in a packet, is denoted by
 $m^{(i)}_p \in \{0,...,M-1\}$. The corresponding modulated chirp is obtained by left-shifting the raw chirp  of $\tau^{(i)}_ p = m^{(i)}_pT  $ in the time domain, yielding a shift of $\frac{m^{(i)}_p}{T_s}$ in the frequency domain (see Fig. \ref{fig:1}):
\begin{align}\label{eq:1} 
s_p^{(i)}(t)=\begin{cases}
&\exp\big(2\jmath\pi\big[\frac{B}{2T_s}t^2+\frac{m_p^{(i)}}{T_s}t\big]\big),\\ 
& \hspace{2.5cm} t \in \text{\big[$-\frac{T_s}{2}, \frac{T_s}{2} - \tau^{(i)}_ p$}\big[,\\
&\exp\big(2\jmath\pi\big[\frac{B}{2T_s}t^2+\big(\frac{m_p^{(i)}}{T_s}-B\big)t\big]\big),\\ 
& \hspace{2.5cm}   t \in \text{ \big[$\frac{T_s}{2} - \tau^{(i)}_ p, \frac{T_s}{2}$}\big[.
\end{cases} 
\end{align}

Then, the complex envelope of the transmitted signal of user $i$, $s^{(i)}(t)$, is given by: 
\begin{equation}
s^{(i)}(t) = \sum _{p=0}^{P-1} s_p^{(i)}(t-pT_s).
\label{eq:sig_and_notation}
\end{equation}

The transmitted LoRa packet is finally obtained by adding the preamble consisting of  $N_p$ raw chirps.
\normalsize 


The rest of this section deals with the reception of a LoRa frame in the presence of interfering users. Indeed, when two users simultaneously transmit their data using the same spreading factor and frequency band,
a collision can possibly occur at the reception. Without loss of generality, only one interfering user is considered. Complete expressions with $N_u$ interferers are directly obtained by adding all interfering terms, which all have a similar expression. 

\begin{figure}[htbp]
    \centering
    \includegraphics[scale=0.5]{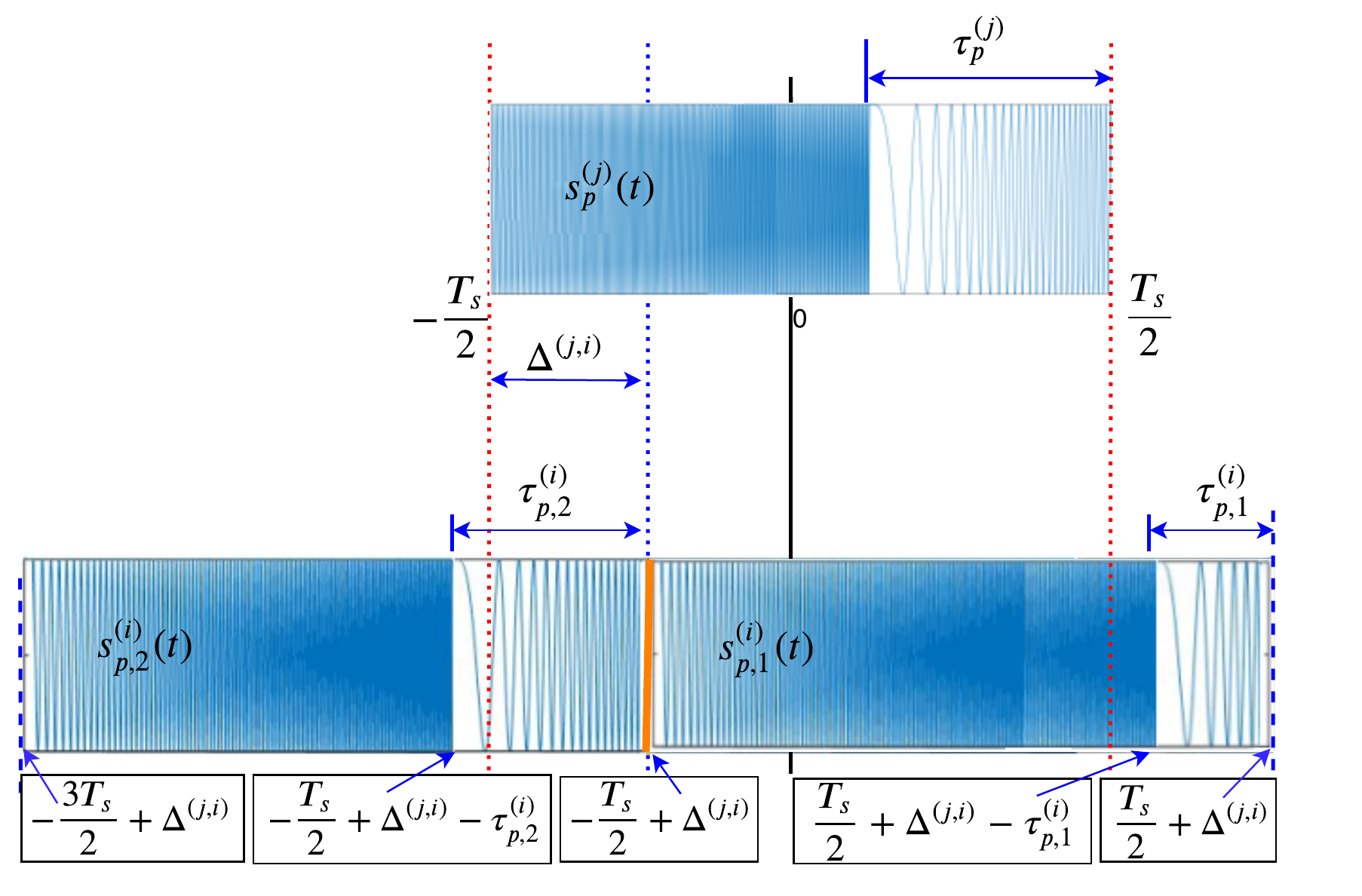}
    \caption{LoRa-like Interfering symbols.}
    \label{fig:2}
\end{figure}

We describe now the collision  between a $p$th symbol of a user of interest $j$ and two consecutive symbols of an interfering user $i$ as shown in Fig. \ref{fig:2}. Recall that $s^{(j)}_p(t)$ denotes the modulated symbol (or chirp) at time $pT_s$ of user $j$ (see (\ref{eq:1})). The two consecutive interfering symbols of user $i$ are denoted by $s_{p,1}^{(j,i)}(t)$ and $s_{p,2}^{(j,i)}(t)$, the index $j$ indicating that the synchronization is carried out on user $j$.
The shifts carrying the corresponding modulating symbols are given by $\tau^{(i)}_{p,1}=m_{p,1}^{(i)}T$ and $\tau^{(i)}_{p,2}=m_{p,2}^{(i)}T$. The delay between user $j$ and user $i$ (accounting for asynchronous transmissions) 
is denoted by $\Delta^{(j,i)}$. 
To formulate the mathematical expression for the interfering signals in Fig. \ref{fig:2}, we consider two conditions that define the positions of the shifts of the interfering symbols:
\setcounter{equation}{5}
\begin{align}\label{eq:5}
\ \text{$C_1$:}\;
\tau^{(i)}_{p,1} >\Delta ^{(j,i)},  
 \text{$C_2$:}\;\;\; 
  \tau^{(i)}_{p,2} <  \Delta ^{(j,i)}. 
\end{align}
\normalsize
Depending on these conditions, there could be two to four 
contributions to the interference.

The expression for the interfering signal is given in \eqref{eq:3} and \eqref{eq:4}. The second case in \eqref{eq:3} (respectively the first in \eqref{eq:4}) contributes only if $C_1$ (respectively  $C_2$) in \eqref{eq:5} is satisfied. 
\normalsize
To synchronize, we perform a correlation between the received signal and the preamble. 
The highest peak indicates the strongest user, indexed by $j$, that we will try to decode first, so we synchronize on it.  

The received signal associated with symbol $p$ of user $j$ sampled at $t=nT$, $n=-\frac{M}{2}, \cdots , \frac{M}{2} - 1$ is:
\begin{align} \label{eq:9}
y_p^{(j)}[n] = h^{(j)}&\sqrt{P_t}\;s_p^{(j)}[n]\nonumber\\
&+h^{(i)}\sqrt{P_t}\left(s_{p,1}^{(j,i)}[n] + s_{p,2}^{(j,i)}[n]\right) + w_p[n].
\end{align}

where we have used the notation $s[n] = s(nT_s)$ for any modulated signal $s(t)$.
$w_p[n] \sim \mathcal{CN}(0,\,\sigma_n^{2})$ is a complex Gaussian noise, $h^{(j)}$ and $h^{(i)}$ are the channel attenuations for users $j$ and $i$, $P_t$ is the transmitted power (we consider the same transmit power for each user).
\begin{figure*}
\hspace{0.1cm}\rule{17.75cm}{0.005cm}
\begin{align}\label{eq:7}
r_p[n]  & = h^{(j)} \sqrt{P_t}\;e^{2\jmath\pi\frac{m_p^{(j)}}{M}n} \nonumber \\
 & + h^{(i)}\sqrt{P_t}\bigg(\sum _{l=1}^{2} e^{2\jmath\pi \phi_{p,l}^{(j,i)}} e^{2\jmath\pi\big(\frac{\varphi_{p,l}^{(j,i)}}{M}\big)n}\mathds{1}_{A_{p,l}}[n] + e^{2\jmath\pi(\phi_{p,l}^{(j,i)}+B\Delta ^{(j,i)})} e^{2\jmath\pi\big(\frac{\varphi_{p,l}^{(j,i)}}{M}-(-1)^{l+1}\big)n}\mathds{1}_{B_{p,l}}[n]\bigg) + w_p[n], \end{align}
\begin{equation} \label{eq:8}
R'_p[k]  = P^{(j)}\delta[k- m_p^{(j)}] + P^{(i)}\Bigg( \sum_{l=1}^{2} e^{\jmath\beta^{(j,i)}_{p,l}} \frac{\sin \big(\pi\big(\frac{k-\varphi^{(j,i)}_{p,l}}{M}\big)\Delta A_{p,l} \big)}{\sin\big(\pi\big(\frac{k-\varphi^{(j,i)}_{p,l}}{M}\big)\big)} + e^{\jmath\gamma^{(j,i)}_{p,l}} \frac{\sin \big(\pi\big(\frac{k-\varphi^{(j,i)}_{p,l}}{M}-(-1)^l\big)\Delta B_{p,l} }{\sin\big(\pi\big(\frac{k-\varphi^{(j,i)}_{p,l}}{M}-(-1)^l\big)\big)} \Bigg) + W'_p[k].
\end{equation}
\hspace{0.5cm}
\normalsize
\begin{align*}
&\text{where}\;\beta^{(j,i)}_{p,l}=2\pi\phi_{p,l}^{(j,i)} -\pi\big(\frac{k-\varphi^{(j,i)}_{p,l}}{M}\big)(\Delta A_{p,l}-1),\; \gamma^{(j,i)}_{p,l}= 2\pi(\phi_{p,l}^{(j,i)} + B\Delta ^{(j,i)})-\pi\Big(\frac{k-\varphi^{(j,i)}_{p,l}}{M}-(-1)^l\Big)(\Delta B_{p,l}-1),\\ 
&\varphi_{p,l}^{(j,i)}= {m_{p,l}^{(i)}-B\Delta ^{(j,i)}}, \;\Delta A_{p,l},\;\Delta B_{p,l}\; \text{are the length of the intervals}\; A_{p,l}\; \text{and}\;B_{p,l}\; \text{for}\;l\in \{1,2\}.
\end{align*}
\hspace{0.1cm}\rule{17.75cm}{0.005cm}
\end{figure*}

To demodulate, the samples of the received signal 
are multiplied by the conjugate of the raw chirp: 
$r_p[n]=y_p^{(j)}[n]c^*[n]$, where $c^*[n]=c^*(nT)$. The resulting signal is given in \eqref{eq:7}, with $A_{p,l}$ and $B_{p,l}$, $l\in\{1,2\}$, from \eqref{eq:3} and \eqref{eq:4}.

\begin{figure}
    \centering
    \includegraphics[scale=0.6]{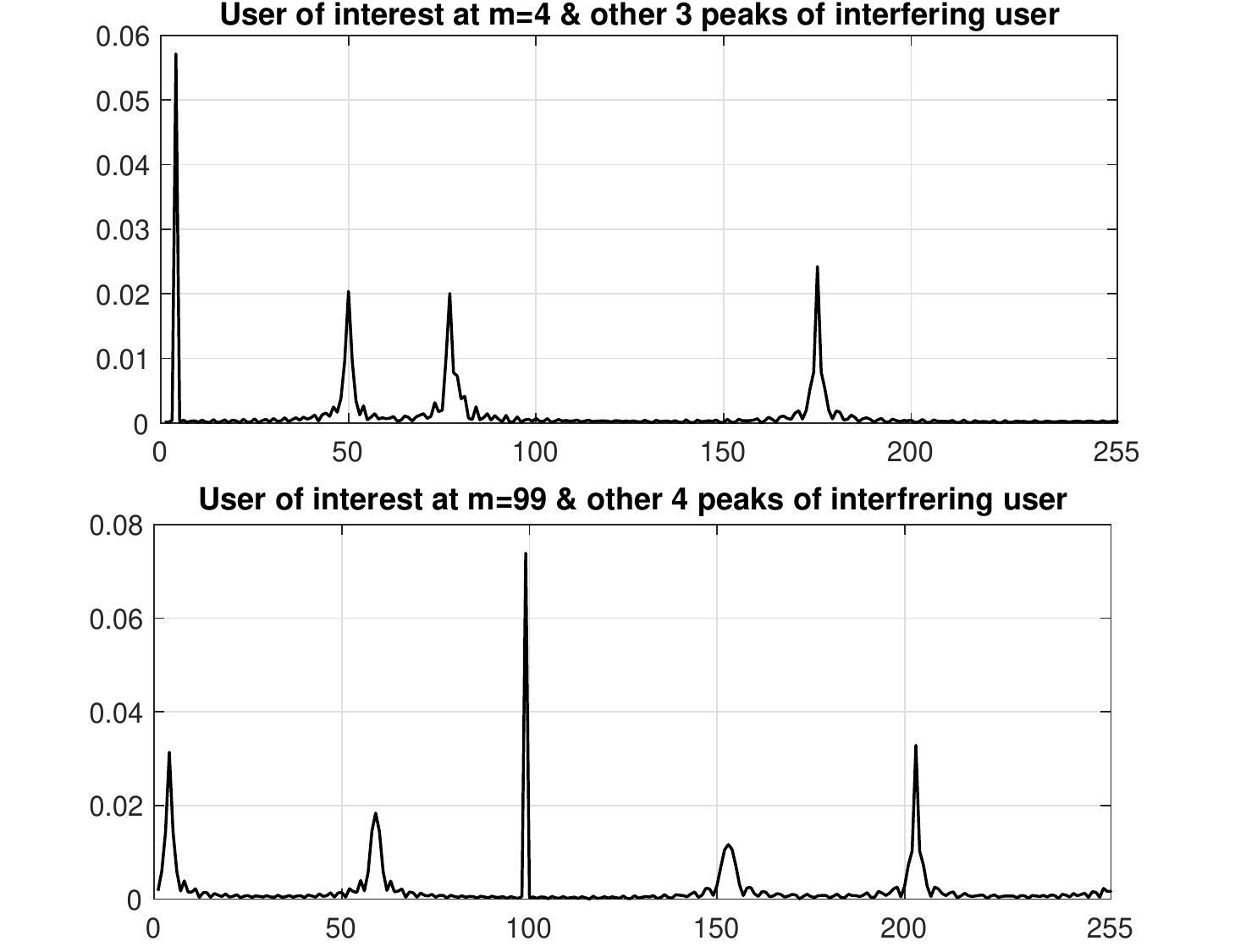}
    \caption{ FFT of coded chirp of the user of interest at the values of 4 and 99 in the presence of one interfering user using $\textrm{SF}= 8$,  $B= 250$ KHz.}
    \label{fig:3}
\end{figure}

Taking the absolute value of the Fast Fourier Transform (FFT) of \eqref{eq:7}, 
the useful information creates a peak whose amplitude depends on the channel attenuation.
Eq. \eqref{eq:7} also shows that user $i$ can create up to four peaks in the Fourier domain (and this is the same for any interfering user). The height of these peaks is determined not only by the channel attenuation but also by the length of the overlaps, and this is good news for our SIC scheme.  
Fig. \ref{fig:3} illustrates this phenomenon by showing the results of the FFT processing for two coded chirps with $m=4$ and $m=99$ 
for the user of interest. The parameters are as follows: Noise level of $-114$ dBm, $\textrm{SF} = 8$, $B= 250$ kHz and one interfering user.
\section{Proposed SIC Receiver and Detection} \label{sec:SIC}
  
We consider SIC as a pure receiver technique, without any power control or any types of adjustment at the transmitter side. 
First, the strongest signal is decoded. Its contribution can then be reconstructed, using an estimate of the channel, 
and subtracted from the received signal. From the resulting residue, 
the second strongest signal can then be extracted \cite{sic}. 
The performance of the decoding will be degraded by two facts: if the difference in received powers from two different users is too close (which means that their channel is similar because we do not perform any kind of power control); and if other peaks collide making the peak of interest not the strongest one and/or significantly deviating from its expected amplitude. It is important to notice in this case that the signal to interference plus noise power ratio, with an interference power evaluated through the sum of all interfering signal powers, is not a reliable metric. We can also note that the wide range of the communications should allow a large dispersion in the received powers, allowing our approach to perform well.  
The decoding of a user can be described as follows.
First, the strongest preamble is detected by calculating the correlation 
between the received signal and the preamble and taking the maximal value, denoted by $C_{\textrm{max}}$.
This allows us to detect the beginning 
of the packet of the strongest user and to estimate its channel $\hat{h}^{(j)}$. 
Multiplication by the down chirp is then performed, see \eqref{eq:7}. 
We apply an FFT to $r_p[n]$ and divide it by the estimated channel, which is shown in {$R'_p[k]={R_p[k]}/{\hat{h}^{(j)}}$, $P^{(j)} =\sqrt{P_t}\;h^{(j)}/\hat{h}^{(j)}$, $P^{(i)} =\sqrt{P_t}\;h^{(i)}/\hat{h}^{(j)}$, $W'_p[k]={W_p[k]}/{\hat{h}^{(j)}}$}, and $W_p[k]$ is the FFT of the Gaussian noise. In \eqref{eq:8}, it is shown that the received power varies according to the standard deviation $\sigma_n$ of the noise term $W'_p[k]$. (Expressions are still given for two users and are simply extended to $N_u$ users.)

The next step is to search for the peaks of the desired user. Instead of the usual maximum search, we propose to search for the peak having the closest value to $P^{(j)}$ by minimizing the difference between the detected peaks in $R'_p[k]$ and $P^{(j)}$. This avoids mistakes when interferers collide and create peaks with a larger amplitude than the desired one.  
We also check that $|R'_p[k]-P^{(j)}|<\varepsilon$ to detect collisions on the desired peak, where $\varepsilon$ is tuned by grid-search, yielding an optimal value of $3\sigma_n$. 
If the above condition is not verified we simply choose the maximum peak. Once the symbols of the desired user is decoded, its contribution is re-constructed and subtracted from the composite received discrete signal $y_p[n]$. The process then starts again to detect the next strongest user. This detection/suppression process is repeated until no more preamble can be found or a packet is not correctly decoded.
Algorithm \ref{alg:receiver} illustrates the proposed receiver. The input 
is only the combined received signal.  
\begin{algorithm}
 \caption{SIC Receiver for LoRa-like Networks}
 \begin{algorithmic}[1]
 \renewcommand{\algorithmicrequire}{\textbf{Input:}}
 \renewcommand{\algorithmicensure}{\textbf{Output:}}
\REQUIRE Received samples $y_p[n]$, number of transmitted symbols $N$, detection threshold $\varepsilon$
\ENSURE  Decoded symbols $m^{(j)}_p$
\STATE $Detected \gets 1$ 
\WHILE {$Detected==1$}
   \IF {$|C_{\textrm{max}}|<\varepsilon$} 
        \STATE $Detected \gets 0$
    \ELSE
        \STATE Estimate $\hat{h}^{(j)}$ 
        \STATE $r_p[n] \gets y_p^{(j)}[n] c^*[n]$
        \STATE $R'_p[k] \gets |FFT(r_p[n])|$  
        \FOR{$p =1 $ to $P$}
            \STATE $u \gets \argminl_k|R'_p[k] - P^{(j)}|$
            \IF {$|R'_p[u] - P^{(j)}|^2< \varepsilon $}
                \STATE $m_p^{(j)} \gets u$ 
            \ELSE
                \STATE $m_p^{(j)} \gets \argmaxl_k(|R'_p[k]^2|)$
            \ENDIF
        \ENDFOR
        \STATE $ \hat{s}_p^{(j)}[n]\gets$ CSS Modulation of $m_p^{(j)}$
        \STATE $y_p[n] \gets y_p[n] - \hat{h}^{(j)}\;\hat{s}_p^{(j)}[n]$
    \ENDIF
     \RETURN $m_p^{(j)}$ 
 \ENDWHILE 
 \end{algorithmic} 
 \label{alg:receiver}
 \end{algorithm}
\section{Simulation Result}
\subsection{Simulation set-up}
In order to evaluate the performance of our proposed scheme, we rely on Monte Carlo simulations of 1000 rounds. 
We consider a circle of radius $R$ with one gateway at the center. Multiple users are uniformly distributed within the circle. The distance from user $i$ to the gateway is denoted by $d^{(i)}$.  
We consider a set of users transmitting their data through block fading channels during the time interval of interest. All users employ CSS modulation with a symbol duration $T_s$ (same SF). They transmit their signals with the same power $P_t$, i.e., there is no power control. We consider asynchronous transmission among the nodes, different devices operating autonomously. 
We consider path loss and Rayleigh multi-path fading $\chi_i$. For path loss, the signal amplitude decays with distance according to ${d^{(i)}}^{-\eta/2}$, where $\eta$ is the path loss exponent. The channel attenuation (in amplitude) is then $h^{(i)}= {d^{(i)}}^{-\eta/2}\chi_i$. In the following we take a maximum range $R=5$ km and we choose $\eta=3.5$. However, the channel attenuation has to be such that the user can be connected to the network with the chosen SF, in other words if their received power is greater than the receiver sensitivity $S$, for instance $S=-124$ dBm for $\textrm{SF}=8$ and $B=250$ kHz. Users that do not respect this condition are discarded and drawn again. The noise level of a receiver at room temperature is $-174 + 10\log_{10}(B) + \textrm{NF}=-114$ dBm, where $\textrm{NF}$ is the receiver noise figure and a classic $6$ dB is considered \cite{LoRa_basic}.
We consider a window of size $\textrm{W}=3T_f$, where $T_f$ is the frame length. $N_u$ devices are transmitting in this window and asynchronicity is ensured through a random variable $\Delta^{(i)}$, uniformly distributed over $\left[ 0,2T_f\right]$. Indeed, $2T_f$ is the vulnerability period of an ALOHA protocol and to be sure we receive all full packets, we need to end our study frame at $3T_f$. A perfect time division would allow 3 users (3 packets) in this frame if collisions are not allowed. 
 
No channel coding is performed and we are interested in the symbol error rate (SER).

\subsection{Performance for a given useful link.}
\begin{figure}
    \centering
    \includegraphics[scale=0.65]{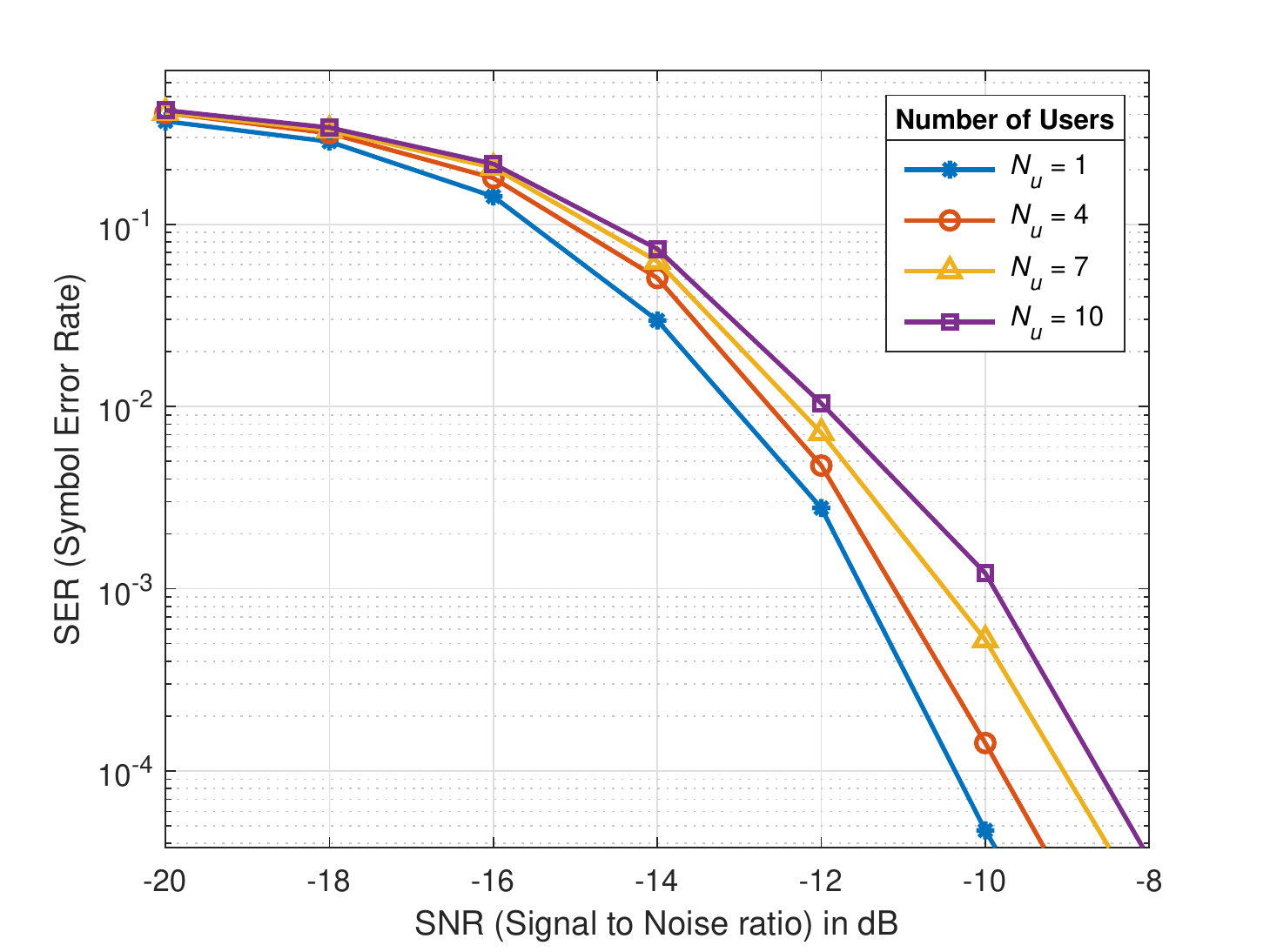}
    \caption{SER for $N_u=1$, $4$, $7$, and $10$, with $\textrm{SF}=8$, $B=250$ kHz, and different SNR values.}
    \label{fig:55}
\end{figure}

For a given useful link characterized by its SNR $\frac{|h|^2P}{\sigma_n^2}$, Fig. \ref{fig:55} shows the performance of this specific receiver when there is no other interfering user ($N_u=1$) as well as when $N_u = 4$, $7$, and $10$ interfering users are present. The positions and channels of these interfering users are randomly chosen at each round. 
For an SNR of $-10$ dB, a typical value for $\textrm{SF}=8$ and $B=250$ kHz, the performance is good even for 10 interfering users, we reach a SER of $10^{-3}$ at SNR of $-10$ dB which is a target SER for LoRa under same SF interference \cite{Analysis_lora}.
\subsection{Mean performance.}

\begin{figure}[htbp]
    \centering
    \includegraphics[scale=0.65]{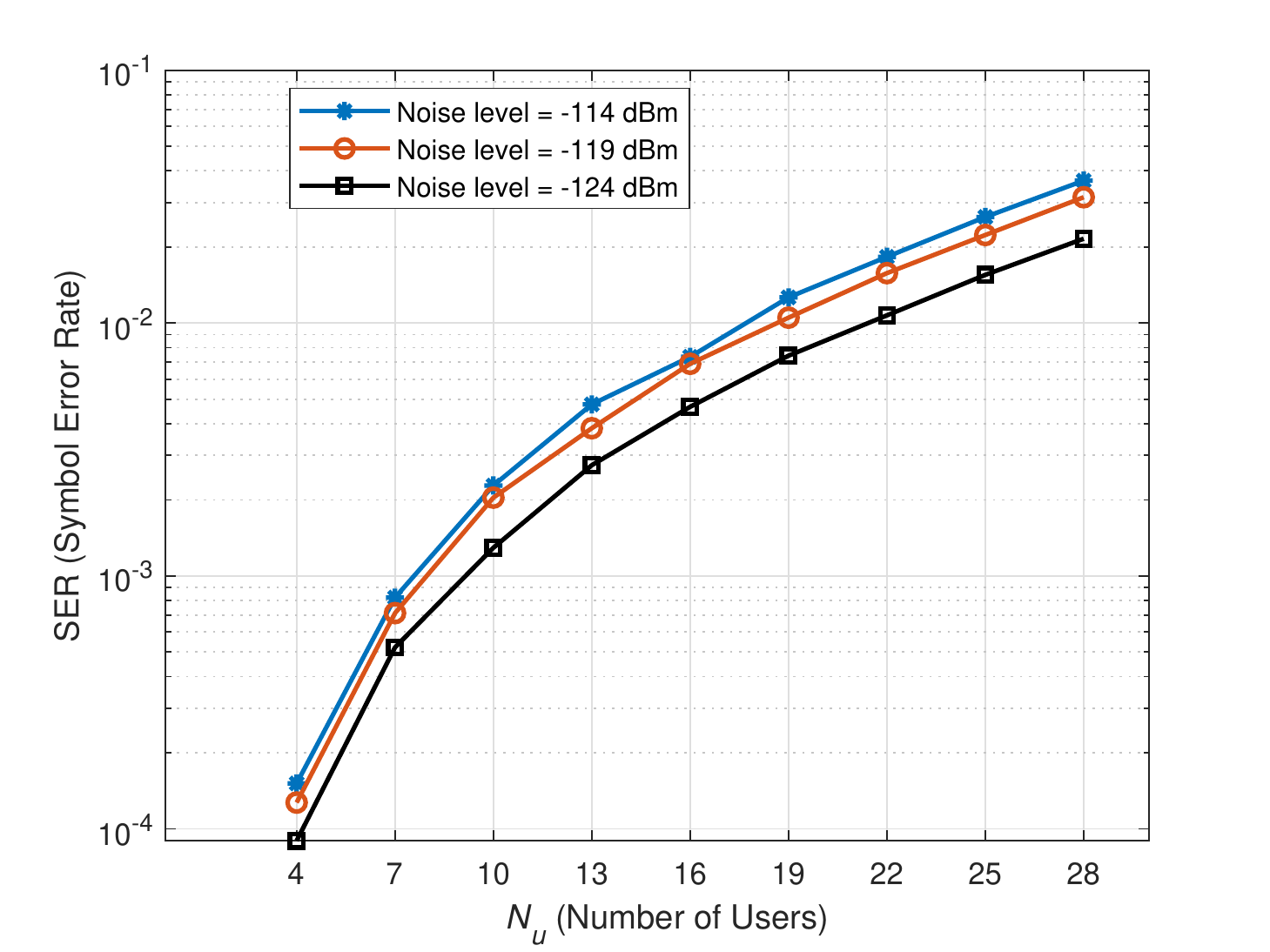}
    \caption{SER for different values of $N_u$ with  $\textrm{SF}=8$, $B=250$ kHz, and several noise levels.}
    \label{fig:4}
\end{figure}

We now evaluate the performance for any user, what ever their channel (so their SNR) is. Fig. \ref{fig:4} shows the average symbol error rate of the proposed receiver with $\textrm{SF}=8$ when the position of all users are randomly drawn at each round and when different noise levels are considered. 
It can be seen that instead of the three users that could be handled with a perfect time division multiple access (TDMA), which is even more than what an ALOHA protocol as the one used in LoRa can handle, a SIC approach allows to have about 20 users with a SER under $10^{-2}$.
\begin{figure}[htbp]
    \centering
    \includegraphics[scale=0.65]{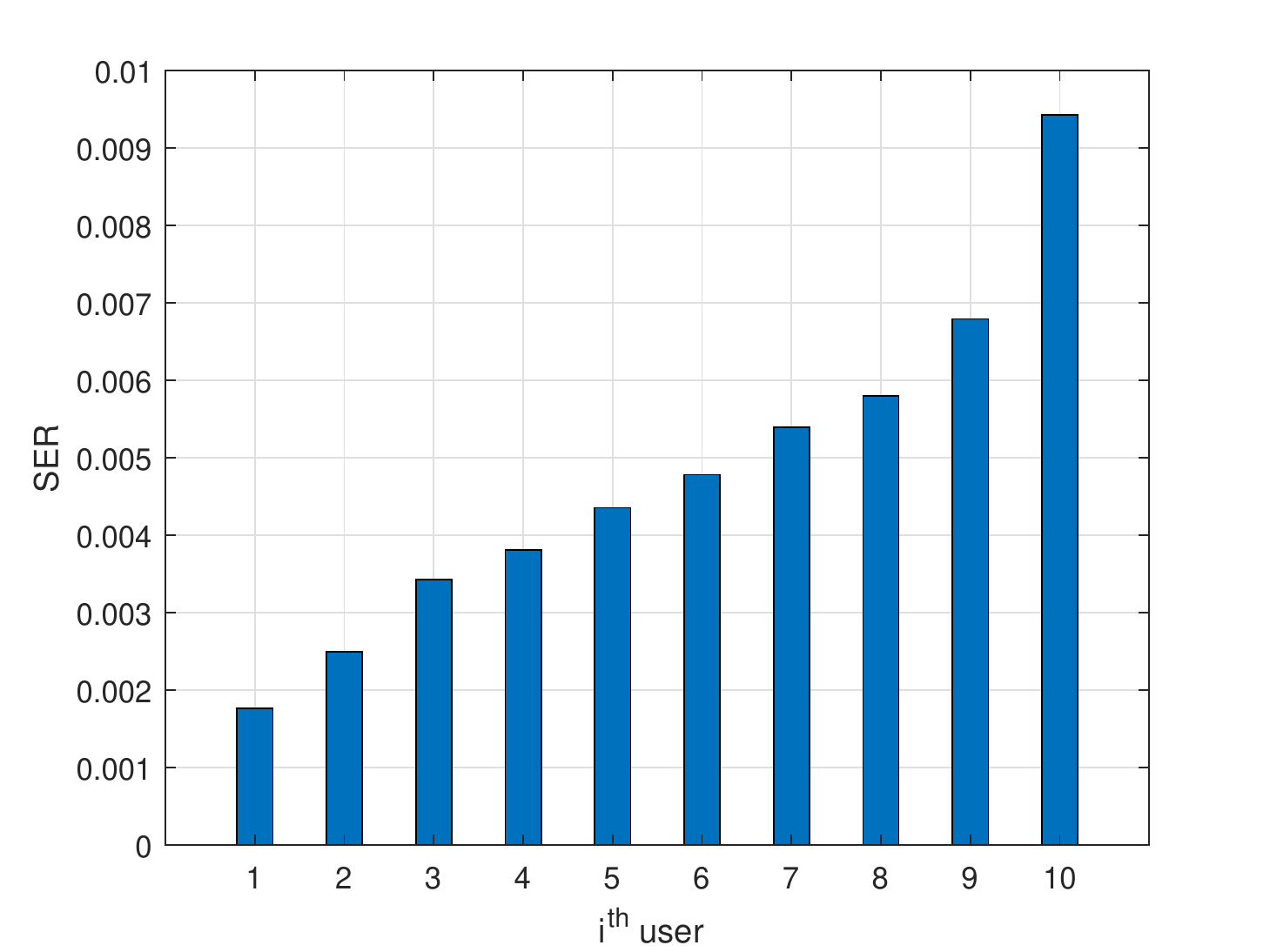}
    \caption{SER of each users for $N_u=10$, $\textrm{SF}=8$, $B=250$ kHz, Noise level of $-114$ dBm.}
    \label{fig:5}
\end{figure}

SIC receiver can be affected by error propagation, consequence of the serial decoding, reconstruction and suppression process. The results is a higher SER for users with lower power. Fig. \ref{fig:5} illustrates the SER of each users when $N_u=10$. 
The users are ordered according to their received power, from the highest (1) to the lowest (10), confirming the error propagation. 

\subsection{\textrm{SF} orthogonality.}
\begin{figure}[htbp]
    \centering
    \includegraphics[scale=0.65]{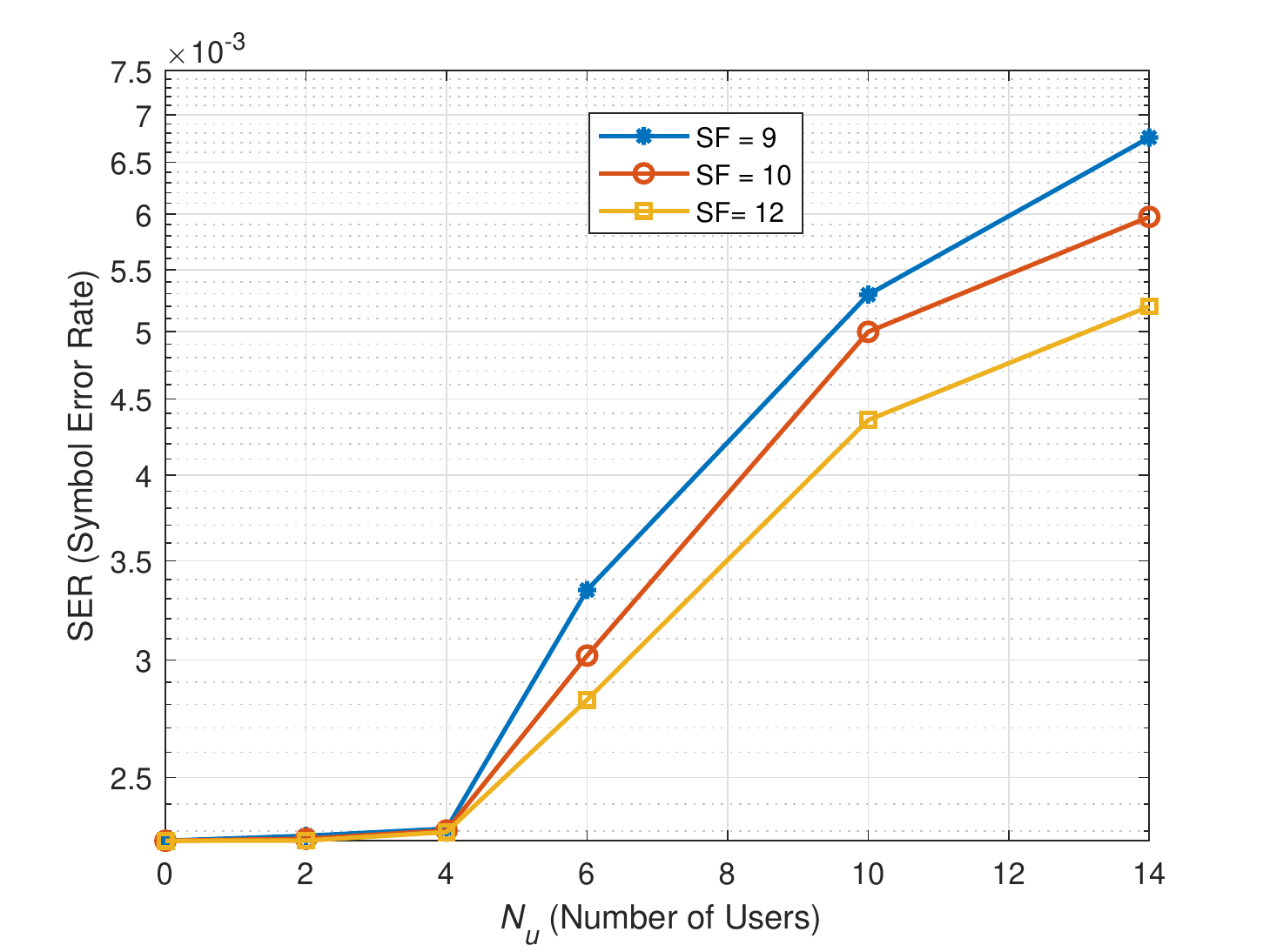}
     \caption{SER of Fixed $10$ users using $\textrm{SF}=8$ and additional interfering users using different SF (9, 10, 12), $B=250$ kHz, Noise level of $-114$ dBm.}
    \label{fig:6}
\end{figure}

It was shown in \cite{LoRa_Imperfect_Orth} that SFs have imperfect orthogonality. Fig. \ref{fig:6} presents the average SER for 10 users, which transmit their signal using $\textrm{SF}=8$. We consider an additional number $N_u$ of interfering users that transmit their signal using a different SF ($9, 10$ or $12$). As long as these added 
interferers are less than four, their impact is negligible. However, the spreading factors are not perfectly orthogonal so that the performance of the 10 users using $\textrm{SF}=8$ degrades when the number of interfering users with a different SF gets larger. We also notice that the shorter SF (9) impacts more than the longer ones. 

\section{Conclusion}
Scalability is a major concern in IoT networks and actual deployed solutions cannot face it. 
However, in the uplink communication, due to the star topology of LPWANs, the complexity can be added at the receiver to improve the system performance, without modifying the low consumption transmission scheme.
And this allows us to propose a serial interference cancellation scheme for chirp spread spectrum based physical layer. Assuming uncoordinated transmitting devices, we use a SIC from the power domain NOMA idea to recover multiple signals arriving simultaneously, on the same bandwidth and with the same SF. The diversity in received powers is ensured by the random radio channels faced by the devices. A SIC receiver allows then to recover the multiple signals.

In fact, though the simulation results we showed that the proposed solution allows supporting 20 times more devices without any modification of the transmission scheme.
This is attained with a first simple implementation of the SIC and further improvement can certainly be obtained with a judicious choice of the PHY layer parameters.

\section*{Acknowledgement}
  This work was supported by IRCICA, USR CNRS 3380, Lille, by the COST action CA15104, IRACON.

\ifCLASSOPTIONcaptionsoff
  \newpage
\fi

\end{document}